\documentclass{svmult}
\usepackage{amssymb}
\usepackage{graphicx}

\newcommand{\eqref}[1]{(\ref{#1})}
\newcommand{\dontprint}[1]{\relax}
\newcommand{\bbR}{{\mathbb{R}}}
\newcommand{\Map}{\mathop{\mathrm{Map}}\nolimits}
\newcommand{\Div}{\mathop{\mathrm{div}}\nolimits}
\newcommand{\tr}{\mathop{\mathrm{tr}}\nolimits}
\newcommand{\Tr}{\mathop{\mathrm{Tr}}\nolimits}

\begin{document}
\title*{Effective Batalin--Vilkovisky theories, equivariant configuration spaces and cyclic chains}
\titlerunning{Effective BV theory and cyclic chains}
\author{Alberto S. Cattaneo\inst{1}\and
Giovanni Felder\inst{2}}

\institute{ Institut f\"ur Mathematik, Universit\"at
Z\"urich-Irchel, Winterthurerstrasse 190, CH-8057 Z\"urich,
Switzerland \texttt{alberto.cattaneo@math.uzh.ch} \and Department of
mathematics, ETH Zurich, CH-8092 Zurich, Switzerland
\texttt{giovanni.felder@math.ethz.ch} }

 \maketitle

\begin{abstract}
Kontsevich's formality theorem states that the differential graded
Lie algebra of multidifferential operators on a manifold $M$ is
$L_\infty$-quasi-isomorphic to its cohomology. The construction of
the $L_\infty$-map is given in terms of integrals of differential
forms on configuration spaces of points in the upper half-plane.
Here we consider configuration spaces of points in the disk and work
equivariantly with respect to the rotation group. This leads to
considering the differential graded Lie algebra of multivector
fields endowed with a divergence operator. In the case of $\mathbb
R^d$ with standard volume form, we obtain an $L_\infty$-morphism of
modules over this differential graded Lie algebra from cyclic chains
of the algebra of functions to multivector fields. As a first
application we give a construction of traces on algebras of
functions with star-products associated with unimodular Poisson
structures. The construction is based on the Batalin--Vilkovisky
quantization of the Poisson sigma model on the disk and in
particular on the treatment of its zero modes.
\end{abstract}

 {\it Dedicated to Murray Gerstenhaber and Jim
Stasheff}
\section{Introduction}\label{s-1}
The Hochschild complex of any algebra with unit carries a
differential graded Lie algebra structure introduced by Gerstenhaber
\cite{Gerstenhaber1963}. In the case of the algebra of smooth
functions on a manifold, one has a differential graded Lie
subalgebra $\mathfrak g_G$ of multidifferential operators, whose
cohomology is the graded Lie algebra $\mathfrak g_S$ of multivector
fields with Schouten--Nijenhuis bracket.\footnote{We use Tsygan's
notation \cite{Tsygan1999}. Kontsevich's notation
\cite{Kontsevich2003} is $\mathcal D_{\mathrm{poly}}=\mathfrak g_G$,
$\mathcal T_{\mathrm{poly}}=\mathfrak g_S$} Kontsevich
\cite{Kontsevich2003} showed that $\mathfrak g_G$ and $\mathfrak
g_S$ are quasi-isomorphic as $L_\infty$-algebras, a notion
introduced by Stasheff as the Lie version of $A_\infty$-algebras
\cite{Stasheff1963}, see \cite{SchlessingerStasheff1985,
LadaStasheff1993}. A striking application of this result is the
classification of formal associative deformations of the product of
functions in terms of Poisson structures. Kontsevich's
$L_\infty$-quasi-isomorphism is given in terms of integrals over
configuration spaces of points in the upper half-plane.  As shown in
\cite{CattaneoFelderPSM}, these are Feynman amplitudes of a
topological quantum field theory known as the Poisson sigma model
\cite{Ikeda,SchallerStrobl}.

In this paper we consider the case of a manifold $M$ endowed with a
volume form $\Omega$. In this case $\mathfrak g_S$ comes with a
differential, the divergence operator $\Div_\Omega$ of degree $-1$.
One considers then the differential graded Lie algebra $\mathfrak
g_S^\Omega=\mathfrak g_S[v]$ where $v$ is an indeterminate of degree
$2$, the bracket is extended by $v$-linearity and the differential
is $v\Div_\Omega$. The relevant topological quantum field theory is
a BF theory  (or Poisson sigma model with trivial Poisson structure)
on a disk whose differential is the Cartan differential on
$S^1$-equivariant differential forms. This theory is described in
Section \ref{s-BV}. The new feature, compared to the original
setting of Kontsevich's formality theorem, is that zero modes are
present. We use recent ideas of Losev, Costello and Mnev to treat
them in the Batalin--Vilkovisky quantization scheme. This gives the
physical setting from which the Feynman amplitudes are derived. In
the remaining sections of this paper, which can be read
independently of Section \ref{s-BV}, we give a purely mathematical
treatment of the same objects. The basic result is the construction
for $M=\mathbb R^d$ of an $L_\infty$-morphism of $\mathfrak
g_S^\Omega$-modules from the module of negative cyclic chains
$(C_{-\bullet}(A)[u], b+uB)$ to the trivial module
$(\Gamma(\wedge^{-\bullet}\mathit{TM}),\Div_\Omega$). We also check
that this $L_\infty$-morphism has properties needed to extend the
result to general manifolds.

As in the case of Kontsevich's theorem, the coefficients of the
$L_\infty$-morphism are integrals of differential forms on
configuration spaces. Whereas Kontsevich considers the spaces of
$n$-tuples of points in the upper half-plane modulo the action of
the group of dilations and horizontal translations, we consider the
space of $n$-tuples of points in the unit disk and work
equivariantly with respect to the action of the rotation group. The
quadratic identities defining the $L_\infty$-relations are then
proved by means of an equivariant version of the Stokes theorem.

As a first application we construct traces in deformation
quantization associated with unimodular Poisson structures. Our
construction can also be extended to the case of supermanifolds; the
trace is then replaced by a nondegenerate cyclic cocycle
(Calabi--Yau structure, see \cite{KontsevichSoibelman}, Section
10.2, and \cite{CostelloCY}) for the $A_\infty$-algebra obtained by
deformation quantization in \cite{CattaneoFelderCoisotropic}.
Further applications will be studied in a separate publication
\cite{CFW}. In particular we will derive the existence of an
$L_\infty$-quasi-isomorphism of $\mathfrak g_S^\Omega$-modules from
the complex $\mathfrak g_S^\Omega$ with the adjoint action to the
complex of cyclic cochains with a suitable module structure. This is
a module version of the Kontsevich--Shoikhet formality conjecture
for cyclic cochains \cite{Shoikhet1999}.

\subsection*{Acknowledgements}
We are grateful to Francesco Bonechi, Damien Calaque, Kevin
Costello, Florian Sch\"atz, Carlo Rossi, Jim Stasheff, Thomas
Willwacher and Marco Zambon for useful comments. This work been
partially supported by SNF Grants 20-113439 and 200020-105450, by
the European Union through the FP6 Marie Curie RTN ENIGMA (contract
number MRTN-CT-2004-5652), and by the European Science Foundation
through the MISGAM program. The first author is grateful to the
Erwin Schr\"odinger Institute, where part of this work was done, for
hospitality.

\subsection*{Notations and conventions}\label{ss-1.1} All vector
spaces are over $\mathbb R$. We denote by $S_n$ the group of
permutations of $n$ letters and by $\epsilon\colon S_n\to \{\pm 1\}$
the sign character. We write $|\alpha|$ for the degree of a
homogeneous element $\alpha$ of a $\mathbb Z$-graded vector space.
The sign rules for tensor products of graded vector spaces hold: if
$f$ and $g$ are linear maps on graded vector spaces, $(f\otimes
g)(v\otimes w)=(-1)^{|g|\cdot|v|}f(v)\otimes g(w)$. The graded
vector space $V[n]$ is $V$ shifted by $n$: $V[n]^i=V^{n+i}$. There
is a canonical map (the identity) $s^n\colon V[n]\to V$ of degree
$n$. The graded symmetric algebra $S^\bullet V=\oplus_{n\geq 0}S^nV$
of a graded vector space $V$ is the algebra generated by $V$ with
relations $a\cdot b=(-1)^{|a|\cdot|b|}b\cdot a$, $a,b\in V$; the
degree of a product of generators is the sum of the degrees. If
$\sigma\in S_n$ is a permutation and $a_1,\dots,a_n\in V$ then
$a_{\sigma(1)}\cdots a_{\sigma(n)}=\epsilon a_1\cdots a_n$; we call
$\epsilon=\epsilon(\sigma;a_1,\dots,a_n)$ the Koszul sign of
$\sigma$ and $a_i$. The exterior algebra $\bigwedge V$ is defined by
the relations $a\wedge b=-(-1)^{|a|\cdot|b|}b\wedge a$ on
generators. We have a linear isomorphism
$S^n(V[1])\to(\wedge^nV)[n]$ given by $v_1\cdots v_n\mapsto
s^{-n}(-1)^{\sum(n-j)(|v_j|-1)}sv_1\wedge\cdots\wedge sv_n$, $v_j\in
V[1]$.

\section{BV formalism and zero modes}\label{s-BV}
This section provides the interested reader with some ``physical''
motivation for the constructions in this paper. It may be safely
skipped by the reader who is only interested in the construction and
not in its motivation.

The basic idea is to use the Batalin--Vilkovisky (BV) formalism in
order to deal with theories with symmetries (like the Poisson sigma
model). What is interesting for this paper is the case when ``zero
modes'' are present.

It is well known in algebraic topology that structures may be
induced on subcomplexes (in particular, on an embedding of the
cohomology) like, e.g., induced differentials in spectral sequences
or Massey products. It is also well known in physics that low-energy
effective field theories may be induced by integrating out
high-energy degrees of freedom. As observed by Losev \cite{Losev}
(and further developed by Mnev \cite{Mnev} and Costello
\cite{Costello}), the two things are actually related in terms of
the BV approach to (topological) field theories. We are interested
in the limiting case when the low-energy fields are just the zero
modes, i.e., the critical points of the action functional modulo its
symmetries.

\newcommand{\calM}{\mathcal{M}}
\newcommand{\calL}{\mathcal{L}}

Let $\calM$ be an SP-manifold, i.e., a graded manifold endowed with
a symplectic form of degree $-1$ and a compatible Berezinian \cite{Schwarz}. Let
$\Delta$ be the corresponding BV-Laplace operator. The compatibility
amounts to saying that $\Delta$ squares to zero and that it
generates the BV bracket $(\ ,\ )$ (i.e., the Poisson bracket of
degree $1$ determined by the symplectic structure of degree $-1$):
namely,
\begin{equation}\label{e:DeltaAB}
\Delta (AB) = (\Delta A)B +(-1)^{|A|} A\Delta B -(-1)^{|A|} (A,B). 
\end{equation}

Assume now that $\calM$ is actually a product of SP-manifolds
$\calM_1$ and $\calM_2$, with BV-Laplace operators $\Delta_1$ and
$\Delta_2$, $\Delta=\Delta_1+\Delta_2$. The central observation is
that for every Lagrangian submanifold $\calL$ of $\calM_2$ and any
function $F$ on $\calM$---for which the integral makes sense---one
has
\begin{equation}\label{e:intDelta}
\Delta_1\int_\calL F = \int_\calL \Delta F.
\end{equation}

In infinite dimensions, where we would really like to work, this
formula is very formal as both the integration and $\Delta$ are
ill-defined. In finite dimensions, on the other hand, this is just a
simple generalization of the fact that, for any differential form
$\alpha$ on the Cartesian product of two manifolds $M_1$ and $M_2$
and any closed submanifold $S$ of $M_2$ on which the integral of
$\alpha$ converges, we have
\[
\D\int_S\alpha = \pm\int_S \D\alpha,
\]
where integration on $S$ yields a differential form on $M_1$. The
correspondence with the BV language is obtained by taking
$\calM_{1,2}:=T^*[-1]M_{1,2}$ and $\calL:=N^*[-1]S$ (where $N^*$
denotes the conormal bundle). The Berezinian on $\calM$ is
determined by a volume form $v= v_1\wedge v_2$ on $M:=M_1\times
M_2$, with $v_i$ a volume form on $M_i$. Finally, $\Delta$ is
$\phi_v^{-1}\circ \D\circ \phi_v$, with
$\phi_v\colon\Gamma(\wedge^\bullet \mathit{TM})\to \Omega^{\dim
M-\bullet}(M)$,
$X\mapsto\phi_v(X):=\iota_X v$.  
The generalization consists in the fact that there are Lagrangian
submanifolds of $\calM_2$ not of the form of a conormal bundle;
however, by a result of Schwarz \cite{Schwarz}, they can always be
brought to this form by a symplectomorphism so that formula
\eqref{e:intDelta} holds in general.

\newcommand{\braket}[2]{\left\langle{\,{#1}\,,\,{#2}\,}\right\rangle}
\newcommand{\calV}{\mathcal{V}}
\newcommand{\sfA}{{\mathsf{A}}}
\newcommand{\sfB}{{\mathsf{B}}}
\newcommand{\calH}{\mathcal{H}}
\newcommand{\sfa}{{\mathsf{a}}}
\newcommand{\sfb}{{\mathsf{b}}}
\newcommand{\sfS}{{\mathsf{S}}}

In the application we have in mind, $\calM_2$ (and so $\calM$) is
infinite-dimensional, but $\calM_1$ is not. Thus, we have a well
defined BV-Laplace operator $\Delta_1$ and try to make sense of
$\Delta$ by imposing \eqref{e:intDelta}, following ideas of
\cite{Losev,Mnev} and, in particular, \cite{Costello}. More
precisely, we consider ``$BF$- like'' theories. Namely, let
$(\calV,\delta)$ and $(\tilde\calV,\delta)$ be complexes with a
nondegenerate pairing $\braket{\ }{\ }$ of degree $-1$ which relates
the two differentials:
\begin{equation}\label{e:braketBA}
\braket\sfB{\delta\sfA}=\braket{\delta\sfB}\sfA,
\qquad\forall\sfA\in\calV,\ \sfB\in\tilde\calV.
\end{equation}
We set $\calM=\calV\oplus\tilde\calV$ and define $S\in
C^\infty(\calM)$ as
\begin{equation}\label{e:action}
S(\sfA,\sfB):=\braket\sfB{\delta\sfA}.
\end{equation}
The pairing defines a symplectic structure of degree $-1$ on $\calM$
and the BV bracket with $S$ is $\delta$. In particular,
\begin{equation}\label{e:ME}
(S,S)=0.
\end{equation}
We denote by $\calH$ ($\tilde \calH$) the
$\delta$-cohomology of $\calV$ ($\tilde\calV)$. Then we
choose an embedding of
$\calM_1:=\calH\oplus\tilde\calH$ into $\calM$ and a complement $\calM_2$.

\begin{example}\label{exa:d}
Take $\calV=\Omega(\Sigma)[1]$ and
$\tilde\calV=\Omega(\Sigma)[s-2]$, with $\Sigma$ a closed, compact
$s$-manifold, and $\delta=\D$, the de~Rham differential, on $\calV$;
up to a sign, $\delta$ on $\tilde\calV$ is also the de~Rham
differential if the pairing is defined by integration:
$\braket\sfB\sfA:=\int_\Sigma\sfB\wedge\sfA$, $\sfA\in\calV$,
$\sfB\in\tilde\calV$. In this case $\calM_1=H(\Sigma)[1]\oplus
H(\Sigma)[s-2]$, with $H(\Sigma)$ the usual de~Rham cohomology. A
slightly more general situation occurs when $\Sigma$ has a boundary;
in this case, appropriate boundary conditions have to be chosen so
that $\delta$ has an adjoint as in \eqref{e:braketBA}. Let
$\partial\Sigma=\partial_1\Sigma \sqcup \partial_2\Sigma$ (each of
the boundary components $\partial_{1,2}\Sigma$ may be empty). We
then choose $\calV=\Omega(\Sigma,\partial_1\Sigma)[1]$ and
$\tilde\calV=\Omega(\Sigma,\partial_2 \Sigma)[s-2]$, where
$\Omega(\Sigma,\partial_i \Sigma)$ denotes differential forms whose
restrictions to $\partial_i \Sigma$ vanish. In this case,
$\calM_1=H(\Sigma,\partial_1\Sigma)[1]\oplus H(\Sigma,\partial_2
\Sigma)[s-2]$.
\end{example}
\begin{example}\label{exa:du}
Suppose that $S^1$ acts on $\Sigma$ (and that the
$\partial_i\Sigma$s are invariant). Let $\vec v$ denote the vector
field on $\Sigma$ generating the infinitesimal action. Let
$\Omega_{S^1}(\Sigma,\partial\Sigma):=\Omega(\Sigma,\partial\Sigma)^{S^1}[u]$
denote the Cartan complex with differential
$\D_{S^1}=\D-u\iota_{\vec{v}}$, where $u$ is an indeterminate of
degree $2$. Then we may generalize Example~\ref{exa:d} replacing
$\Omega(\Sigma,\partial\Sigma)$ with
$\Omega_{S^1}(\Sigma,\partial\Sigma)$.
\end{example}

\newcommand{\vevo}[1]{{\left\langle\;{#1}\;\right\rangle}_0}
\newcommand{\calO}{\mathcal{O}}

Now suppose that $\calH$ (and so $\tilde\calH$) is
finite-dimensional, as in the examples above. In this case it is
always possible to choose a BV-Laplacian $\Delta_1$ on $\calM_1$.
Once and for all we also choose a Lagrangian submanifold $\calL$ on
which the infinite-dimensional integral makes sense in perturbation
theory. Assuming $\Delta S=0$, the first consequence of
\eqref{e:intDelta} and \eqref{e:ME} is that the partition function
\[
Z_0 = \int_\calL \E^{\frac \I\hbar S}
\]
is $\Delta_1$-closed. Actually, in the case at hand, $Z_0$ is constant on $\calM_1$.

For every functional $\calO$ on $\calM$ for which integration on
$\calL$ makes sense, we define the expectation value
\[
\vevo\calO:=\frac{\int_\calL \E^{\frac \I\hbar S}\,\calO}{Z_0}
\]
The second consequence of \eqref{e:intDelta}, and of the fact that
$Z_0$ is constant on $\calM_1$, is the Ward identity
\begin{equation}\label{e:Ward}
\Delta_1\vevo\calO = \vevo{\Delta\calO - \frac \I\hbar\delta\calO}.
\end{equation}
where we have also used \eqref{e:DeltaAB}.

To interpret the Ward identity for $\calO=\sfB\otimes\sfA$, we
denote by $\{\theta^\mu\}$ a linear coordinate system on $\calH$ and
by $\{\zeta_\mu\}$ a linear coordinate system on $\tilde\calH$, such
that their union is a Darboux system for the symplectic form on
$\calM_1$ with
$\Delta_1=\frac{\partial}{\partial\theta^\mu}\frac\partial{\partial\zeta_\mu}$.
We next write $\sfA=\alpha_\mu\theta^\mu+\sfa$ and
$\sfB=\beta^\mu\zeta_\mu+\sfb$ with $\sfa\oplus\sfb\in\calM_2$.
The left hand side 
of the Ward identity is now simply $\Delta_1\vevo{\sfB\otimes\sfA}=\sum_\mu(-1)^{|\beta^\mu|}\beta^\mu\otimes\alpha_\mu=:\phi$. On the assumption
that the ill-defined BV-Laplacian $\Delta$ should be a second order
differential operator, the first term $\vevo{\Delta(\sfB\otimes\sfA)}$ on the right hand side is
ill-defined but constant on $\calM_1$; we denote it by $K$.
Since $\delta$ vanishes in cohomology and, as a differential operator, it can be extracted from the expectation value,
\eqref{e:Ward} yields a constraint for the propagator
\begin{equation}\label{e:prop}
\omega := \frac \I{\hbar} \vevo{\sfb\otimes\sfa};
\end{equation}
namely,
\[
\delta\omega = K - \phi.
\]

{}From now on we assume that $\calM$ is defined in terms of
differential forms as in Examples~\ref{exa:d} and~\ref{exa:du}. In
this case, $\omega$ is a distributional $(s-1)$-form on
$\Sigma\times\Sigma$ while $\phi$ is a representative of the
Poincar\'e dual of the diagonal $D_\Sigma$ in $\Sigma\times\Sigma$.
By the usual naive definition of $\Delta$, $K$ is equal to the delta
distribution  on $D_\Sigma$. Thus, the restriction of $\omega$ to the
configuration space $C_2(\Sigma):= \Sigma\times\Sigma\setminus
D_\Sigma$ is a smooth $(m-1)$-form satisfying $\D\omega=\phi$. If
$\Sigma$ has a boundary, $\omega$ satisfies in addition the
conditions $\iota_1^*\omega=\iota_2^*\omega=0$ with $\iota_1$ the
inclusion of $\Sigma\times\partial_1\Sigma$ into
$\Sigma\times\Sigma$ and $\iota_2$ the inclusion of
$\partial_2\Sigma\times\Sigma$ into $\Sigma\times\Sigma$. Denoting
by $\pi_{1,2}$ the two projections $\Sigma\times\Sigma\to\Sigma$ and
by $\pi^{1,2}_*$ the corresponding fiber-integrations, we may define
$P\colon\Omega(\Sigma,\partial_1\Sigma)\to\Omega(\Sigma,\partial_1\Sigma)$
and $\tilde
P\colon\Omega(\Sigma,\partial_2\Sigma)\to\Omega(\Sigma,\partial_2\Sigma)$
by $P(\sigma)=\pi^2_*(\omega\wedge\pi_1^*\sigma)$ and $\tilde
P(\sigma)=\pi^1_*(\omega\wedge\pi_2^*\sigma)$. Then the equation for
$\omega$ implies that $P$ and $\tilde P$ are parametrices for the
complexes $\Omega(\Sigma,\partial_1 \Sigma)$ and
$\Omega(\Sigma,\partial_2 \Sigma)$; namely,
$\D P+P\D = 1-\varpi$ and $\D \tilde P+\tilde P\D = 1-\tilde\varpi$, where $\varpi$ and $\tilde\varpi$ denote the projections onto cohomology.  

This characterization of the propagator of a ``$BF$-like'' theory
also appears in \cite{Costello}. Even though not justified in terms
of the BV formalism, this choice of propagator was done before in
\cite{BottCattaneo} for Chern--Simons theory out of purely
topological reasons, and later extended to $BF$ theories in
\cite{CattaneoRossi}. A propagator with these properties also
appears in \cite{Ferrario} for the Poisson sigma model on the
interior of a polygon.

The quadratic action \eqref{e:action} is usually the starting point
for a perturbative expansion. The first singularity that may occur
comes from evaluating $\omega$ on $D_\Sigma$ (``tadpole''). A mild
form of renormalization consists in removing tadpoles or, in other
words, in imposing that $\omega$ should vanish on $D_\Sigma$. By
consistency, one has then to set $K$ equal to the restriction of
$\phi$ to $D_\Sigma$. In other words, one has to impose
\begin{equation}\label{e:Deltareg}
\Delta(\sfB(x)\sfA(x))=\psi(x):=\sum_\mu(-1)^{|\beta^\mu|}\beta^\mu(x)\alpha_\mu(x),\qquad\forall
x\in\Sigma.
\end{equation}
Observe that $\psi$ is a representative of the Euler class of
$\Sigma$. By \eqref{e:DeltaAB} and \eqref{e:Deltareg} one then
obtains a well-defined version of $\Delta$ on the algebra $C^\infty(\calM)'$
generated
by local functionals. This may be regarded as an asymptotic version
(for the energy scale going to zero) of Costello's regularized
BV-Laplacian \cite{Costello}.
Actually,
\begin{lemma}
$(C^\infty(\calM)',\Delta)$ is a BV algebra.
\end{lemma}

We now restrict ourselves to the setting of the Poisson sigma model
\cite{Ikeda,SchallerStrobl}. Namely, we assume $\Sigma$ to be
two-dimensional and take
$\calV=\Omega(\Sigma,\partial_1\Sigma)[1]\otimes(\bbR^m)^*$ and
$\tilde\calV=\Omega(\Sigma,\partial_2 \Sigma)\otimes\bbR^m$. Here
$(\bbR^m)^*\times\bbR^m$ is a local patch of the cotangent bundle of
an $m$-dimensional target manifold $M$. (Whatever we say here and in
the following may be globalized by taking $\calM$ to be the graded
submanifold of $\Map(T[1]\Sigma,T^*[1]M)$ defined by the given
boundary conditions.) There is a Lie algebra morphism from the
graded Lie algebra $\mathfrak g_S=\Gamma(\wedge^{\bullet+1}
\mathit{TM})$ of multivector fields on $M$ to $C^\infty(\calM)'$
endowed with the BV bracket \cite{CattaneoFelderAKSZ}: to $\gamma\in
\Gamma(\wedge^{k} \mathit{TM})$ it associates the local functional
\[
S_\gamma = \int_\Sigma
\gamma^{i_1,\dots,i_k}(\sfB)\,\sfA_{i_1}\cdots\sfA_{i_k}.
\]
Moreover, for $k>0$, $(S,{S_\gamma})=0$. With the regularized
version of the BV-Laplacian, we get
\[
\Delta S_\gamma = \int_\Sigma
\psi\,(\Div_\Omega\gamma)^{i_1,\dots,i_{k-1}}(\sfB)\,\sfA_{i_1}\cdots\sfA_{i_{k-1}},
\]
where $\Div_\Omega$ is the divergence with respect to the constant
volume form $\Omega$ on $\bbR^n$. To account for this
systematically, we introduce the differential graded Lie algebra
$\mathfrak g_S^\Omega:=\mathfrak g_S[v]$, where $v$ is an
indeterminate of degree two and the differential $\delta_\Omega$ is
defined as $v\Div_\Omega$ (and the Lie bracket is extended by
$v$-linearity). To $\gamma\in\Gamma(\wedge^{k} \mathit{TM})\,v^l$ we
associate the local functional
\[
S_\gamma = (-\I\hbar)^l\int_\Sigma
\psi^l\,\gamma^{i_1,\dots,i_k}(\sfB)\,\sfA_{i_1}\cdots\sfA_{i_k}.
\]
It is now not difficult to prove the following
\begin{lemma}
The map $\gamma\mapsto S_\gamma$ is a morphism of differential
graded Lie algebras from $(\mathfrak g_S^\Omega,[\ ,\
],\delta_\Omega)$ to $(C^\infty(\calM)',(\ ,\ ),-\I\hbar\Delta)$.
Moreover, for every $\gamma\in\Gamma(\wedge^{k} \mathit{TM})\,v^l$
with $k$ or $l$ strictly positive, we have $(S,{S_\gamma})=0$. If
$\partial\Sigma=\emptyset$, the last statement holds also for
$k=l=0$.
\end{lemma}
Observe that  $\psi^2=0$ by dimensional reasons. However,
in the generalization to the equivariant setting of Example~\ref{exa:du}, higher powers of $\psi$ survive.

A first application of this formalism is the Poisson sigma model on
$\Sigma$. If $\pi$ is a Poisson bivector field (i.e.,
$\pi\in\Gamma(\wedge^2\mathit{TM})$, $[\pi,\pi]=0$), then
$\sfS_\pi:=S+S_\pi$ satisfies the master equation
$(\sfS_\pi,\sfS_\pi)=0$ but in general not the quantum master
equation $\frac12(\sfS_\pi,\sfS_\pi)+\I\hbar\Delta\sfS_\pi=0$, which
by \eqref{e:DeltaAB} is equivalent to $\Delta
\E^{\frac\I\hbar\sfS_\pi}=0$. Unless $\psi$ is trivial\footnote{If
the class of $\psi$ is trivial, one may always choose bases in the
embedded cohomologies so that $\psi=0$. If one does not want to make
this choice, one observes anyway that for $\psi=\D\tau$ one has
$-\I\hbar\Delta S_\gamma=(S,S'_\gamma)$ with $S'_\gamma =
(-\I\hbar)^l\int_\Sigma
\tau\psi^{l-1}\,\gamma^{i_1,\dots,i_k}(\sfB)\,\sfA_{i_1}\cdots\sfA_{i_k}$,
for $\gamma\in\Gamma(\wedge^{k} \mathit{TM})\,v^l$, $l>0$. In the
case at hand, one may then define a solution of the quantum master
equation as $S+S_\pi+S'_\pi$.} (which is, e.g., the case for
$\Sigma$ the upper half plane, as in \cite{CattaneoFelderPSM}, or
the torus), this actually happens only if $\pi$ is divergence free.
More generally, if $\pi$ is unimodular \cite{Koszul}, by definition
we may find a function $f$ such that $\Div_\Omega \pi=[\pi,f]$. So
$\tilde\pi:=\pi+vf$ is a Maurer--Cartan element in $\mathfrak
g_S^\Omega$ (i.e.,
$\delta_\Omega\tilde\pi-\frac12[\tilde\pi,\tilde\pi]=0$). Hence
$\sfS_{\tilde\pi}:=S+S_{\tilde\pi}$ satisfies the quantum master
equation. It is not difficult to check that, for $\psi$ nontrivial,
the unimodularity of $\pi$ is a necessary and sufficient condition
for having a solution of the quantum master equation of the form
$S+S_\pi + O(\hbar)$. For $\Sigma$ the sphere this was already
observed in \cite{BonechiZabzine} though using slightly different
arguments.

We will now restrict ourselves to the case of interest for the rest
of the paper: namely, $\Sigma$ the disk and
$\partial_2\Sigma=\emptyset$. In this case $H(\Sigma)$ is
one-dimensional and concentrated in degree $0$ while
$H(\Sigma,\partial\Sigma)$ is one-dimensional and concentrated in
degree two. Thus, $\calH=(\bbR^m)^*[-1]$ and $\calH=\bbR^m$ which
implies $\calM_1=T^*[-1]M$. Functions on $\calM_1$ are then
multivector fields on $M$ but with reversed degree and the operator
$\Delta_1$ turns out to be the usual divergence operator
$\Div_\Omega$ (which is now of degree $+1$) for the constant volume
form. A first simple application is the expectation value
\[
\tr g:=\frac{\int_\calL \E^{\frac \I\hbar
\sfS_{\tilde\pi}}\,\calO_g}{Z_0}=\vevo{\E^{\frac \I\hbar
S_{\tilde\pi}}\,\calO_g},\qquad g\in C^\infty(M),
\]
where $\tilde\pi$ is a Maurer--Cartan element corresponding to a
unimodular Poisson structure and $O_g(\sfA,\sfB):=g(\sfB(1))$, with
$1$ in $\partial\Sigma$ which we identify with the unit circle.
Consider now $\tr_2(g,h):=\vevo{\E^{\frac \I\hbar
S_{\tilde\pi}}\,\calO_{g,h}}$, with
$\calO_{g,h}:=g(\sfB(1))\int_{\partial\Sigma\setminus\{1\}}h(\sfB)$.
By \eqref{e:DeltaAB}, we then have
$\Delta_1\tr_2(g,h)=\vevo{\E^{\frac \I\hbar
S_{\tilde\pi}}\,\delta\calO_{g,h}}$. Arguing as in
\cite{CattaneoFelderPSM}, we see that the right hand side
corresponds to moving the two functions $g$ and $h$ close to each
other (in the two possible ways) and by ``bubbling'' the disk around
them; so we get
\[
\Delta_1\tr_2(g,h)=\tr{g\star h}-\tr{h\star g},
\]
where $\star$ is Kontsevich's star product \cite{Kontsevich2003}
which corresponds to the Poisson sigma model on the upper half plane
\cite{CattaneoFelderPSM}. Since $\Delta_1$ is just the divergence
operator with respect to the constant volume form $\Omega$, for
compactly supported functions we have the trace
\[
\Tr g:=\int_M \tr g\;\Omega.
\]
More generally, we may work out the Ward identities relative to the
quadratic action \eqref{e:action} (there is also an equivariant
version for $S^1$ acting by rotations on $\Sigma$). Given
$a_0,a_1,\dots,a_p$ in $C^\infty(M)$ (or in $C^\infty(M)[u]$ for the
equivariant version), we define
\[
\calO_{a_0,\dots,a_p}:=
a_0(\sfB(1))\int_{t_1<t_2<\dots<t_p\in\partial\Sigma\setminus
\{1\}}a_1(\sfB)\cdots a_p(\sfB)
\]
and
\[
G_n(\gamma_1,\dots,\gamma_n;a_0,\dots,a_p):=\vevo{S_{\gamma_1}\dots
S_{\gamma_n}\,\calO_{a_0,\dots,a_p}},
\]
$\gamma_i\in\mathfrak g_S^\Omega$, $i=1,\dots,n$. By \eqref{e:Ward}
we then have
\begin{eqnarray*}
-\I\hbar\Delta_1G_n(\gamma_1,\dots,\gamma_n;a_0,\dots,a_p)&=&
-\I\hbar \vevo{\Delta(S_{\gamma_1}\dots S_{\gamma_n}\,\calO_{a_0,\dots,a_p})} + \\
&\phantom{=}&+\vevo{\delta(S_{\gamma_1}\dots
S_{\gamma_n}\,\calO_{a_0,\dots,a_p})}.
\end{eqnarray*}
The left hand side is just $(-\I\hbar)$ times the divergence
operator applied to the multivector field $G_n$. The first term on
the right hand side may then be computed as
\begin{eqnarray*}
-\I\hbar \vevo{\Delta(S_{\gamma_1}\dots S_{\gamma_n}\,\calO_{a_0,\dots,a_p})} =\\
=\sum_{i=1}^n (-1)^{\sigma_i} G_n(\gamma_1,\dots,{\delta_\Omega\gamma_i},\dots,\gamma_n;a_0,\dots,a_p) +\\
-\I\hbar\sum_{1\le i<j\le n}  (-1)^{\sigma_{ij}}
G_{n-1}([\gamma_i,\gamma_j],\gamma_1,\dots,\hat\gamma_i,\dots,\hat\gamma_j,\dots,\gamma_n;a_0,\dots,a_p),
\end{eqnarray*}
where the caret denotes omission and
\begin{eqnarray*}
\sigma_i &:= &{\sum_{c=1}^{i-1}|\gamma_c|},\\
\sigma_{ij} &:=
&{|\gamma_i|\sum_{c=1}^{i-1}|\gamma_c|+|\gamma_j|\sum_{c=1,\,c\not=i}^{j-1}|\gamma_c|+|\gamma_i|+1}.
\end{eqnarray*}
with $|\gamma|=k$ for $\gamma\in\Gamma(\wedge^k\mathit{TM})[v]$. The
second term on the right hand side is a boundary contribution (in
the equivariant sense if $\delta=\D_{S^1}=\D-u\iota_{\vec{v}}$). By
bubbling as in \cite{CattaneoFelderPSM}, some of the $\gamma_i$s
collapse together with some of the consecutive $a_k$s and the
result---which is Kontsevich's formality map---is put back into $G$.
The whole formula can then be interpreted as an $L_\infty$-morphism
from the cyclic Hochschild complex to the complex of multivector
fields regarded as $L_\infty$-modules over $\mathfrak g_S^\Omega$,
as we are going to explain in the rest of the paper.

The only final remark is that $\I\hbar$ occurs in this formula only
as a book keeping device. We define $F_n$ by formally setting
$\I\hbar=1$ in $G_n$. 



\section{Hochschild chains and cochains of algebras of smooth functions}\label{s-2}

Kontsevich's theorem states that there is an
$L_\infty$-quasi-isomorphism from the graded Lie algebra $\mathfrak
g_S=\Gamma(\wedge^{\bullet+1} \mathit{TM})$ of multivector fields on
a smooth manifold $M$, with the Schouten--Nijenhuis bracket and
trivial differential, to the differential graded Lie algebra
$\mathfrak g_G$ of multidifferential operators on $M$ with
Gerstenhaber bracket and Hochschild differential. Through
Kontsevich's morphism the Hochschild and cyclic chains become a
module over $\mathfrak g_S$. In this section we review these notions
as well as results and conjectures about them.
\subsection{Multivector fields and multidifferential operators}
 \label{ss-2.1}
Let $\mathfrak g_S$ be the graded vector space $\mathfrak
g_S=\oplus_{j\geq{-1}}\mathfrak g^j_S$ of multivector fields:
$\mathfrak g^{-1}_S=C^\infty(M), \mathfrak
g_S^0=\Gamma(\mathit{TM}), \mathfrak
g_S^{1}=\Gamma(\wedge^2\mathit{TM})$, and so on. The
Schouten--Nijenhuis bracket of multivector fields is defined to be
the usual Lie bracket on vector fields and is extended to arbitrary
multivector field by the Leibniz rule:
$[\alpha\wedge\beta,\gamma]=\alpha\wedge[\beta,\gamma]+
(-1)^{|\gamma|\cdot(|\beta|+1)}[\alpha,\gamma]\wedge\beta,$
$\alpha,\beta,\gamma\in\mathfrak g_S$. The graded Lie algebra
$\mathfrak g_S$ is considered here as a differential graded Lie
algebra with trivial differential.

The differential graded Lie algebra $\mathfrak g_G$ of
multidifferential operators is, as a complex, the subcomplex of the
shifted Hochschild complex $\mathrm{Hom}(A^{\otimes(\bullet+1)},A)$
of the algebra $A=C^\infty(M)$ of smooth functions, consisting of
multilinear maps that are differential operators in each argument.
The Gerstenhaber bracket \cite{Gerstenhaber1963} on $\mathfrak g_G$
is the graded Lie bracket
$[\phi,\psi]=\phi\bullet_G\psi-(-1)^{|\phi|\cdot|\psi|}
\psi\bullet_G\phi$ with Gerstenhaber product\footnote{The sign
differs by a factor $(-1)^{|\phi|\cdot|\psi|}$ from the sign in
\cite{Gerstenhaber1963}. We have chosen the convention making the
induced bracket on cohomology equal to the standard
Schouten--Nijenhuis bracket on multivector fields}
\begin{equation}\label{e-Gerstenhaberproduct}
\phi\bullet_G\psi=\sum_{k=0}^n(-1)^{|\psi|(|\phi|-k)}\phi\circ(
\mathrm{id}^{\otimes k}\otimes\psi\otimes\mathrm{id}^{\otimes
|\phi|-k}).
\end{equation}
The Hochschild differential can be written in terms of the bracket
as $[\mu,\cdot]$, where $\mu\in \mathfrak
g_{G}^1=\mathrm{Hom}(A\otimes A,A)$ is the multiplication in $A$.

The Hochschild--Kostant--Rosenberg map $\mathfrak
g_S^\bullet\to\mathfrak g_G^\bullet$ induces an isomorphism of
graded Lie algebras on cohomology. It is the identity on $\mathfrak
g_S^{-1}=C^\infty(M)=\mathfrak g_G^{-1}$ and, for any vector fields
$\xi_1,\dots,\xi_n$, it sends the multivector field
$\xi_1\wedge\dots\wedge\xi_n$ to the multidifferential operator
\[
 f_1\otimes\cdots \otimes f_n \mapsto
 \frac1{n!}\sum_{\sigma\in S_n}
 \epsilon(\sigma)
 \xi_{\sigma(1)}(f_1)\cdots\xi_{\sigma(n)}(f_n),\quad f_i\in A.
\]
Although the HKR map is a chain map inducing a Lie algebra
isomorphism on cohomology, it does not respect the Lie bracket at
the level of complexes. The correct point of view on this problem
was provided by Kontsevich in his formality conjecture, which he
then proved in \cite{Kontsevich2003}. The differential graded Lie
algebras $\mathfrak g_S$, $\mathfrak g_G$ should be considered as
$L_\infty$-algebras and the HKR map is the first component of an
$L_\infty$-morphism. Let us recall the definitions.
\subsection{$L_\infty$-algebras}\label{ss-2.2}
For any graded vector space $V$ let $S^+V=\oplus_{j=1}^\infty S^jV$
be the free coalgebra without counit cogenerated by $V$. The
coproduct is $\Delta(a_1\cdots a_n)=\sum_{p=1}^{n-1}\sum_\sigma \pm
a_{\sigma(1)}\cdots a_{\sigma(p)}\otimes a_{\sigma(p+1)}\cdots
a_{\sigma(n)}$, with summation over shuffle permutations with Koszul
signs. A coderivation of a coalgebra is an endomorphism $D$ obeying
$\Delta\circ D=(D\otimes \mathrm{id}+\mathrm{id}\otimes D)\circ
\Delta$. Coderivations with the commutator bracket form a Lie
algebra. What is special about the free coalgebra $S^+V$ is that for
any linear map $D\colon S^+V\to V$ there is a unique coderivation
$\tilde D$ such that $D=\pi\circ \tilde D$, where $\pi$ is the
projection onto $V=S^1V$. By definition an $L_\infty$-algebra is a
graded vector space $\mathfrak g$ together with a coderivation $D$
of degree 1 of $S^+(\mathfrak g[1])$ obeying $[D,D]=0$. A
coderivation is thus given by a sequence of maps (the Taylor
components) $D_n\colon S^n\mathfrak{g}[1]\to \mathfrak{g}[2]$ (or
$\wedge^n\mathfrak g\to\mathfrak g[2-n]$), $n=1,2,\dots$, obeying
quadratic relations. In particular $D_1$ is a differential and $D_2$
is a chain map obeying the Jacobi identity up to a homotopy $D_3$.
It follows that $D_2$ induces a Lie bracket on the $D_1$-cohomology.
Differential graded Lie algebras are $L_\infty$-algebras with
$D_3=D_4=\cdots=0$. An $L_\infty$-morphism $(\mathfrak
g,D)\rightsquigarrow (\mathfrak g', D')$ is a homomorphisms $U\colon
S^+\mathfrak g[1]\to S^+\mathfrak g'[1]$ of graded coalgebras such
that $U\circ D=D'\circ U$. Homomorphisms of free coalgebras are
uniquely defined by their composition with the projection
$\pi'\colon S^+\mathfrak{g}'[1]\to \mathfrak{g}'[1]$; thus $U$ is
uniquely determined by its Taylor components $U_n\colon S^n\mathfrak
g[1]\to\mathfrak g'[1]$ (or $\wedge^n\mathfrak g\to \mathfrak
g'[1-n]$): $U_n$ is the restriction to $S^n\mathfrak{g}[1]$ of
$\pi'\circ U$. Conversely, any such sequence $U_n$ comes from a
coalgebra homomorphism. The first relation between $D, D'$ and $U$
is that $U_1$ is a chain map.

\begin{theorem}\label{t-ko} (Kontsevich \cite{Kontsevich2003}) There is an
$L_\infty$-morphism $\mathfrak g_S(M)\rightsquigarrow \mathfrak
g_G(M)$ whose first Taylor component $U_1$ is the
Hochschild--Kostant--Rosenberg map.
\end{theorem}
If $M$ is an open subset of $\mathbb R^d$ the formula for the Taylor
components $U_n$ is explicitly given in \cite{Kontsevich2003} as a
sum over Feynman graphs.

\subsection{Multivector fields and differential forms}\label{ss-3.1}
 The algebra $\Omega^\bullet(M)$ of differential forms on a
manifold $M$ is a module over the differential graded Lie algebra
$\mathfrak g_S(M)$ of multivector fields: a multivector field
$\gamma\in\Gamma(\wedge^{p+1}\mathit{TM})$ acts on forms as
$\mathcal L_\gamma\omega=\D\iota_\gamma+(-1)^p\iota_\gamma d$
generalizing Cartan's formula for Lie derivatives of vector fields.
Here $\D$ is the de Rham differential and the interior
multiplication $\iota_\gamma$ is the usual multiplication if
$\gamma$ is a function and is the composition of interior
multiplications of vector fields $\xi_j$ if
$\gamma=\xi_1\wedge\cdots\wedge\xi_k$. Moreover the action of
$\mathfrak g_S(M)$ on $\Omega^\bullet(M)$ commutes with the de Rham
differential and induces the trivial action on cohomology.
\subsection{Hochschild cochains and cyclic chains}\label{ss-3.2}
The algebras $\Omega^\bullet(M)$ and $H^\bullet(M)$ are cohomologies
of the complexes of the Hochschild and of the periodic cyclic chains
of $C^\infty(M)$. The normalized Hochschild chain complex of a
unital algebra $A$ is $C_{\bullet}(A)=A\otimes \bar A^{\otimes
\bullet}$, where $\bar A=A/\mathbb{R}1$. If we denote by
$(a_0,a_1,\dots,a_p)$ the class of $a_0\otimes \cdots\otimes a_p$ in
$C_{p}(A)$, the Hochschild differential is
\begin{eqnarray*}
b(a_0,\dots,a_p)&=&\sum_{i=0}^{p-1}(-1)^i(a_0,\dots,a_ia_{i+1},\dots,a_p)
\\
&&+(-1)^p(a_pa_0,a_1,\dots,a_{p-1}).
\end{eqnarray*}
We set $C_p(A)=0$ for $p<0$. There is an HKR map
$C_{\bullet}(A)\to\Omega^{\bullet}(M)$ given by
\begin{equation}\label{e-HKR}
(a_0,\dots,a_p)\mapsto \frac1{p!}a_0\D a_1\cdots \D a_p.
\end{equation}
It is a chain map if we consider differential forms as a complex
with {\em trivial} differential. The HKR map induces an isomorphism
on homology, provided we take a suitable completion of the tensor
product $C^\infty(M)^{\otimes(p+1)}$, for example the jets at the
diagonal of smooth maps $M^{p+1}\to \mathbb R$.  On the Hochschild
chain complex there is a second differential $B$ of degree $1$ and
anticommuting with $b$, see \cite{Connes1985}:
\[
B(a_0,\dots,a_p)=\sum_{i=0}^{p}(-1)^{ip}(1,a_i,\dots,a_p,a_0,\dots
a_{i-1}).
\]
The negative cyclic complex, in the formulation of
\cite{GetzlerJones}, is $CC^{-}_{-\bullet}(A)=C_{-\bullet}(A)[u]$
with differential $b+uB$, where $u$ is of degree $2$. The extension
of the HKR map by $\mathbb R[u]$-linearity defines a
quasi-isomorphism
\[
(CC^-_{-\bullet}(A),b+uB)\to (\Omega^{-\bullet}(M)[u],u\,d)
\]
Now both $C(A)$ and $CC^-(A)$ are differential graded modules over
the Lie algebra $\mathfrak g_G$ of multidifferential operators. The
action is the restriction of the action of cochains on chains
$C^k(A)\otimes C_{p}(A)\to C_{p-k+1}(A)$, $\phi\otimes a\mapsto
\phi\cdot a$, defined for any associative algebra with unit as
\begin{eqnarray*}
\lefteqn{(-1)^{(k-1)(p+1)}\phi \cdot (a_0,\dots,a_p)}\\
  &=&
 \sum_{i=0}^{p-k+1}
 (-1)^{i(k-1)}
 (a_0,\dots,a_{i-1},
 \phi(a_i,\dots,a_{i+k-1}),
 a_{i+k},\dots,a_p)
 \\
 &&+\sum_{i=p-k+2}^p
 (-1)^{ip}
 (\phi(a_i,\dots,a_p,a_0,\dots,a_{i+k-p-2}),
 a_{i+k-p-1},\dots,a_{i-1}).
\end{eqnarray*}
This action extends by $\mathbb R[u]$-linearity to an action on the
negative cyclic complex.

\subsection{$L_\infty$-modules}\label{ss-3.3}
Let $(\mathfrak g,D)$ be an $L_\infty$-algebra. The free
$S^+\mathfrak g[1]$-comodule generated by a vector space $V$ is
$\hat V =S\mathfrak g[1]\otimes V$ with coaction $\Delta^V\colon
\hat V\to S^+\mathfrak g[1]\otimes \hat V$ defined as
\[
\Delta^V(\gamma_1\cdots\gamma_n\otimes
v)=\sum_{p=1}^{n}\sum_{\sigma\in S_{p,n-p}}\pm
\gamma_{\sigma(1)}\cdots\gamma_{\sigma(p)}\otimes
(\gamma_{\sigma(p+1)}\cdots\gamma_{\sigma(n)}\otimes v).
\]
A coderivation of the $L_\infty$-module $V$ is then an endomorphism
$D^V$ of $\hat V$ obeying $\Delta^V\circ D^V=(D\otimes
\mathrm{id}+\mathrm{id}\otimes D^V)\Delta^V$. An $L_\infty$-module
is a coderivation $D^V$ of degree $1$ of $\hat V$ obeying $D^V\circ
D^V=0$. A coderivation is uniquely determined by its composition
with the projection $\hat V\to V$ onto the first direct summand and
is thus given by its Taylor components $D^V_n\colon S^n\mathfrak
g[1]\otimes V\to  V[1]$. The lowest component $D^V_0$ is then a
differential on $V$ and $D^V_1$ a chain map inducing an honest
action of the Lie algebra $H(\mathfrak g, D_1)$ on the cohomology
$H(V,D_0^V)$. A morphism of $L_\infty$-modules $V_1\to V_2$ over
$\mathfrak g$ is a degree 0 morphism of $S^+\mathfrak
g[1]$-comodules $F\colon\hat V_1\to \hat V_2$ intertwining the
coderivations. The composition with the projection $\hat V_2\to V_2$
gives rise to Taylor components
\[
F_n\colon S^n\mathfrak g[1]\otimes V_1\to V_2,\qquad n=0,1,2,\dots
\]
that determine $F$ completely. The lowest component $F_0$ is then a
chain map inducing a morphism of $H(\mathfrak g,D_1)$-modules on
cohomology.


\subsection{Tsygan and Kontsevich conjectures \cite{Tsygan1999}, \cite{Shoikhet1999}}\label{ss-3.4}
\begin{conjecture}\label{conj1}
 There exists a quasi-isomorphism of $L_\infty$-modules
 \[
F\colon C_{-\bullet}(C^\infty(M))\rightsquigarrow
(\Omega^{-\bullet}(M), 0)
\] such that $F_0$ is the HKR map.
\end{conjecture}
\begin{conjecture}\label{conj2}
There exists a natural $\mathbb C[[u]]$-linear quasi-isomorphism of
$L_\infty$-modules
 \[
 F\colon CC_{-\bullet}^{-} (C^\infty(M))\rightsquigarrow (\Omega^{-\bullet}(M)[[u]], ud)
 \]
  such that
$F_0$ is the Connes quasi-isomorphism \cite{Connes1985}, given by
the $u$-linear extension of the HKR map \eqref{e-HKR}
\end{conjecture}
Conjecture \ref{conj1} is now a theorem. Different proofs for
$M=\mathbb R^d$ were given in \cite{TamarkinTsygan2001} and
\cite{Shoikhet2003}. Shoikhet's proof \cite{Shoikhet2003} gives an
explicit formula for the Taylor components of $F$ in terms of
integrals over configuration spaces on the disk and extends to
general manifolds, as shown in \cite{Dolgushev2006}.

Let us turn to Kontsevich's formality conjecture for cyclic
cochains, as quoted in \cite{Shoikhet1999}. Recall that a volume
form $\Omega\in\Omega^{d}(M)$ on a $d$-dimensional manifold defines
an isomorphism $\Gamma(\wedge^k\mathit{TM})\to \Omega^{d-k}(M)$,
$\gamma\mapsto\iota_\gamma\Omega$. The de Rham differential $\D$ on
$\Omega^\bullet(M)$ translates to a differential $\Div_\Omega$, the
divergence operator of degree $-1$. The divergence operator is a
derivation of the bracket on $\mathfrak
g_S=\Gamma(\wedge^{\bullet+1}\mathit{TM})$ of degree $-1$. Let us
introduce the differential graded Lie algebra $ \mathfrak
g^\Omega_S=(\mathfrak g_S[v],\delta_\Omega), $ where $v$ is a formal
variable of degree $2$. The bracket is the Schouten--Nijenhuis
bracket and the differential is $\delta_\Omega=v\Div_\Omega$. The
cyclic analogue of $\mathfrak g_G$ is the differential graded Lie
algebra
\[
\mathfrak g_G^{\mathrm{cycl}}=\left\{\varphi\in \mathfrak
g_G,\,\int_Ma_0\varphi(a_1,\dots,a_p)\Omega=(-1)^p
\int_Ma_p\varphi(a_0,\dots,a_{p-1})\Omega\right\}.
\]

\begin{conjecture}\label{conj3}
For each volume form $\Omega\in\Omega^d(M)$ there exists an
$L_\infty$-quasi-isomorphism of $L_\infty$-algebras
$
F\colon  \mathfrak g^\Omega_S\rightsquigarrow \mathfrak g^{\mathrm{cycl}}_G.
$
\end{conjecture}

Shoikhet \cite{Shoikhet1999} constructed a quasi-isomorphism of
complexes $C_1:\mathfrak g^\Omega_S\to\mathfrak g^{\mathrm{cycl}}_G$
and conjectural formulae for an $L_\infty$-morphism whose first
component is $C_1$ in terms of integrals over configuration spaces.
One consequence of the conjecture is the construction of
cyclically-invariant star-products from divergenceless Poisson
bivector fields. Such star-products were then constructed
independently of the conjecture, see \cite{FelderShoikhet}.


\section{An $L_\infty$-morphism for cyclic chains}  \label{s-4}
\subsection{The main results}                        \label{ss-4.1}
Let $\Omega$ be volume form on a manifold $M$ and $\mathfrak
g_S^\Omega$ be the differential graded Lie algebra $\mathfrak
g_S[v]$ with Schouten bracket and differential
$\delta_\Omega=v\Div_\Omega$, see Section \ref{ss-3.4}. The
Kontsevich $L_\infty$-morphism composed with the canonical
projection $\mathfrak g_S^\Omega\to\mathfrak g_S=\mathfrak
g_S^\Omega/v\mathfrak g_S^\Omega$ is an $L_\infty$-morphism
$\mathfrak g_S^\Omega\rightsquigarrow \mathfrak g_G$. Through this morphism the
differential graded $\mathfrak g_G$-module $CC_\bullet^-(A)$ of
negative cyclic chains of $A=C^\infty(M)$ becomes an
$L_\infty$-module over $\mathfrak g_S^\Omega$.

\begin{theorem}\label{t-Grace}
Let $M$ be an open subset of $\mathbb R^d$ with coordinates
$x_1,\dots, x_n$ and volume form $\Omega=\D x_1\cdots \D x_d$. Let
$A=C^\infty(M)$. Let $\Gamma(\wedge^{-\bullet}\mathit{TM})$ be the
differential graded module over $\mathfrak g^\Omega_S$ with
differential $\Div_\Omega$ and trivial $\mathfrak
g^\Omega_S$-action. Then there exists an $\mathbb R[u]$-linear
morphism of $L_\infty$-modules over $\mathfrak g_S^\Omega$
\[
F\colon CC_{{-\,\bullet}}^-(A)\rightsquigarrow
\Gamma(\wedge^{-\bullet} \mathit{TM})[u],
\]
such that
 \begin{enumerate}\item[(i)] The component $F_0$  of $F$ vanishes on $CC_p(A)$, $p>0$ and
for $f\in A\subset CC^-_0(A)$, $F_0(f)=f$.
 \item[(ii)] For
$\gamma\in\Gamma(\wedge^k\mathit{TM})$, $\ell=0,1,2,\dots$,
$a=(a_0,\dots,a_p)\in CC^-_p(A)$,
\[
F_1(\gamma v^\ell;a)=\left\{\begin{array}{ll}(-1)^p u^{s}\gamma
\lrcorner H(a),& \mbox{if $k\geq p$ and $s=k+\ell-p-1\geq
0$.}\\
0,&\mbox{otherwise.}
\end{array}
\right.
\]
 Here
$\lrcorner\colon\Gamma(\wedge^k\mathit{TM})\otimes\Omega^p(M)\to\Gamma(\wedge^{k-p}\mathit{TM})$
is the contraction map and $H$ is the HKR map \eqref{e-HKR}.
\item[(iii)] The maps $F_n$ are equivariant under linear coordinate
transformations and $F_n(\gamma_1\cdots\gamma_n;a)=\gamma_1\wedge
F_{n-1}(\gamma_2\cdots\gamma_n;a)$ whenever
$\gamma_1=\sum(c^i_{k}x_k+d^i)\partial_i\in\mathfrak{g}_S\subset\mathfrak{g}_S^\Omega$
is an affine vector field and
$\gamma_2,\dots,\gamma_n\in\mathfrak{g}_S^\Omega$.
\end{enumerate}
\end{theorem}
The proof of this Theorem is deferred to Section~\ref{ss-6.6}.

In explicit terms, $F$ is given by a sequence of $\mathbb
R[u]$-linear maps $F_n\colon S^n\mathfrak g^\Omega_S[1]\otimes
CC^-(A)\to\Gamma(\wedge^n\mathit{TM})$, $\gamma\otimes a\mapsto
F_n(\gamma;a)$, $n\geq0$, obeying the following relations. For any
$\gamma=\gamma_1\cdots\gamma_n\in S^n\mathfrak g^\Omega_S[1]$, $a\in
CC^-_p(A)$.
\begin{eqnarray}\label{e-Miranda}
 && F_n(\delta_\Omega \gamma;a) +(-1)^{|\gamma|+p} F_n(\gamma;(b+uB)a)
  \\
 &+&\sum_{k=0}^{n-1}
 \sum_{\sigma\in S_{k,n-k}}(-1)^{|\gamma|-1}
 \epsilon(\sigma;\gamma) F_k(\gamma_{\sigma(1)}\cdots\gamma_{\sigma(k)};
 U_{n-k}(\bar\gamma_{\sigma(k+1)}\cdots\bar\gamma_{\sigma(n)})\cdot
 a)\nonumber
 \\
 &+&\sum_{i<j}\epsilon_{ij}
 F_{n-1}((-1)^{|\gamma_i|-1}[\gamma_i,\gamma_j]\cdot\gamma_1\cdots
 \hat\gamma_i\cdots\hat\gamma_j\cdots\gamma_n;a) =\Div_\Omega\,
 F_n(\gamma;a).         \nonumber
\end{eqnarray}
Here $\bar \gamma_i$ denotes the projection of $\gamma_i$ to
$\mathfrak g_S[1]=\mathfrak g_S^\Omega[1]/v\mathfrak g_S^\Omega[1]$;
$S_{p,q}\subset S_{p+q}$ is the set of $(p,q)$-shuffles and the
signs $\epsilon(\sigma;\gamma),\epsilon_{ij}$ are the Koszul signs
coming from the permutation of the $\gamma_i\in\mathfrak g_S[1]$;
$|\gamma|=\sum_i|\gamma_i|$; the differential $\delta_\Omega$ is
extended to a degree 1 derivation of the algebra $S\mathfrak
g_S^\Omega[1]$; the maps $U_k\colon S^k\mathfrak g_S[1]\to \mathfrak g_G[1]$
are the Taylor components of the Kontsevich
$L_\infty$-morphism of Theorem \ref{t-ko}.

We give the explicit expressions of the maps  $F_n$ in Section
\ref{s-5}. 
Before that we explore some consequences.
\subsection{Maurer--Cartan elements}
An element of degree 1 in $\mathfrak g_S^\Omega$ has the form
$\tilde\pi=\pi+v h$ where $\pi$ is a bivector field and $h$ is a
function. The Maurer--Cartan equation
$\delta_\Omega\tilde\pi-\frac12[\tilde\pi,\tilde\pi]=0$ translates
to
\[
[\pi,\pi]=0,\qquad \Div_\Omega\,\pi -[h,\pi]=0.
\]
Thus $\pi$ is a Poisson bivector field whose divergence is a
Hamiltonian vector field with Hamiltonian $h$. Such Poisson
structures are called unimodular \cite{Koszul}. As explained in
\cite{Kontsevich2003}, Poisson bivector fields in $\epsilon\mathfrak
g_S[[\epsilon]]$ are mapped to solution of the Maurer--Cartan
equations in $\epsilon\mathfrak g_G[[\epsilon]]$, which are
star-products, i.e., formal associative deformations of the
pointwise product in $C^\infty(M)$:
\[
f\star
g=fg+\sum_{n=1}^\infty\frac{\epsilon^n}{n!}U_n(\pi,\dots,\pi)(f\otimes
g).
\]
Here the function part of $\tilde\pi$ does not contribute as it is
projected away in the $L_\infty$-morphism $\mathfrak
g^\Omega_S\rightsquigarrow \mathfrak g_G$.

If $\tilde\pi=\pi+vh$ is a Maurer--Cartan element in $\mathfrak
g_S^\Omega$ then $\tilde\pi_\epsilon=\epsilon\pi+vh$ is a
Maurer--Cartan element in $\mathfrak g_S^\Omega[[\epsilon]]$. The
twist of $F$ by $\tilde\pi$ then gives a chain map from the
negative cyclic complex of the algebra $A_\epsilon =(
C^\infty(M)[[\epsilon]],\star)$ to $\Gamma(\wedge
\mathit{TM})[u][[\epsilon]]$. In particular we get a trace
\begin{equation}\label{e-Derek}
f\mapsto\tau(f)=\sum_{n=0}^\infty\frac1{n!} \int_M
F_n(\tilde\pi_\epsilon,\dots,\tilde\pi_\epsilon;f)\Omega,
 \end{equation}
on the subalgebra of $A_\epsilon$ consisting of functions with
compact support. Here there is a question of convergence since there
are infinitely many terms contributing to each fixed power of
$\epsilon$. The point is that these infinitely many terms combine to
exponential functions. More precisely we have the following result.
\begin{proposition}\label{p-Mirelle}
The trace \eqref{e-Derek} can be written as
 \[
 \tau(f)=\sum_{n=0}^\infty\frac{\epsilon^n}{n!}\int_M
 H_n(\pi,h,f)\E^h\Omega=\int_Mf\E^h\Omega+O(\epsilon)
 \]
where $H_n$ is a differential polynomial in $\pi,h,f$.
\end{proposition}

The proof is based on the expression of $F_n$ in terms of graphs. We
postpone it to Section \ref{ss-5last}, after we introduce this
formalism.

\section{Feynman graph expansion of the $L_\infty$-morphism}     \label{s-5}
  In this section we construct the morphism of
$L_\infty$-modules of Theorem \ref{t-Grace}. The Taylor components
have the form
\[
F_n(\gamma;a)=\sum_{\Gamma\in \mathcal G_{\mathbf{k},m}}w_\Gamma
F_\Gamma(\gamma;a).
\]
Here $\gamma=\gamma_1\cdots\gamma_n$, with
$\gamma_i\in\Gamma(\wedge^{k_i}\mathit{TM})[v]$,
$\mathbf{k}=(k_1,\dots,k_n)$ and $a=(a_0,\dots,a_m)\in C_m(A)$. The
sum is over a finite set $\mathcal G_{\mathbf k,m}$ of directed
graphs with some additional structure. To each graph a weight
$w_\Gamma\in\mathbb R[u]$, defined as an integral over a
configuration space of points in the unit disk, is assigned.

We turn to the descriptions of the graphs and weights.

\subsection{Graphs}                                 \label{ss-5.1}
Let $m,n\in\mathbb Z_{\geq0}$, $\mathbf{k}=(k_1,\dots,k_n)\in\mathbb
Z_{\geq0}^n$. We consider directed graphs $\Gamma$ with $n+m$
vertices with additional data obeying a set of rules. The data are a
partition of the vertex set into three totally ordered subsets
$V(\Gamma)=V_1(\Gamma)\sqcup V_2(\Gamma)\sqcup V_w(\Gamma)$, a total
ordering of the edges originating at each vertex and the assignment
of a non-negative integer, the \emph{degree}, to each vertex in
$V_1(\Gamma)$. The rules are
\begin{enumerate}
\item
There are $n$ vertices in $V_1(\Gamma)$. There are exactly $k_i$
edges originating at the $i$th vertex of $V_1(\Gamma)$.
\item There are $m$ vertices in $V_2(\Gamma)$.
There are no edges originating at these vertices.
\item There is exactly one edge pointing at each vertex in $V_w(\Gamma)$
and no edge originating from it.
\item There are no edges starting and ending at the same vertex.
\item For each pair of vertices $i,j$ there is at most one edge from
$i$ to $j$.
\end{enumerate}
The last rule is superfluous, but since all graphs with multiple
edges will have vanishing weight we may just as well exclude them
from the start. This has the notational advantage that we may think
of the edge set $E(\Gamma)$ as a subset of $V(\Gamma)\times
V(\Gamma)$.

 Two graphs are called equivalent if there is a graph
isomorphism between them that respects the partition and the
orderings. The set of equivalence classes is denoted $\mathcal
G_{\mathbf{k},m}$.

The vertices in $V_1(\Gamma)$ are called vertices of the first type,
those in $V_2(\Gamma)$ of the second type. The vertices in
$V_b(\Gamma)=V_1(\Gamma)\cup V_2(\Gamma)$ are called black, those in
$V_w(\Gamma)$ are called white. We denote by $E_b(\Gamma)$ the subset
of $E(\Gamma)$ consisting of edges whose endpoints are black.

 To each $\Gamma\in\mathcal G_{\mathbf{k},m}$ there corresponds
a multivector field $F_\Gamma(\gamma;a)$ whose coefficients are
differential polynomials in the components of $\gamma_i,a_i$. The
rules are the same as in \cite{Kontsevich2003} except for the
additional white vertices, representing uncontracted indices and the
degrees $d_i$, that select the power $d_i$ of $v$ in $\gamma_i$. Let
us consider for example the graph of Figure \ref{f-Sheepman} and
suppose that the degrees of the two vertices of the first type are $k$ and
$\ell$. The algebra of multivector fields on $M\subset\mathbb R^d$
is generated by $C^\infty(M)$ and anticommuting generators
$\theta_\nu=\partial/\partial x_\nu$. Thus
$\gamma\in\Gamma(\wedge^k\mathit{TM})$ has the form
 \[ \gamma=\frac 1{k!}\sum_{\nu_1,\dots,\nu_k} \gamma^{\nu_1\dots
 \nu_k}\theta_{\nu_1}\cdots\theta_{\nu_k}.
 \]
 The components $\gamma^{\nu_1\dots \nu_k}\in C^\infty(M)$ are
 skew-symmetric under permutation of the indices $\nu_i$.
The graph of Figure \ref{f-Sheepman}, with the convention that the
edges originating at each vertex are ordered counterclockwise, gives
then
\[
F_\Gamma(\gamma_1v^k,\gamma_2v^\ell;a_0,a_1,a_2)=\sum\gamma_1^{ij}\partial_{j}
\gamma_2^{pqr}\partial_ia_0\partial_pa_1\partial_qa_2\theta_r,
\]
and is zero on other monomials in $v$.
\setlength{\unitlength}{2cm}
\begin{figure}
\centering
\begin{picture}(3,1.5)
 \put(0,0){\circle*{0.1}}
 \put(0.1,-0.05){$a_0$}
 \put(1.2,0){\circle*{0.1}}
 \put(1.3,-0.05){$a_1$}
 \put(2.4,0){\circle*{0.1}}
 \put(2.5,-0.05){$a_2$}
 \put(0,1.2){\circle*{0.1}}
 \put(0,1.35){$\gamma_1v^k$}
 \put(1.2,1.2){\circle*{0.1}}
 \put(1.2,1.35){$\gamma_2v^\ell$}
 \put(2.4,1.2){\circle{0.1}}
 \put(0,1.1){\vector(0,-1){1}}
 \put(0.1,1.2){\vector(1,0){1}}
 \put(1.3,1.1){\vector(1,-1){1}}
 \put(1.3,1.2){\vector(1,0){1}}
 \put(1.2,1.1){\vector(0,-1){1}}
\end{picture}
\caption{A graph in $\mathcal G_{(2,3),3}$ with two vertices in
$V_1$ of valencies $(2,3)$, three in $V_2$ and one white vertex. The
degrees of the vertices of the first type are $k$ and
$\ell$.}\label{f-Sheepman}
\end{figure}
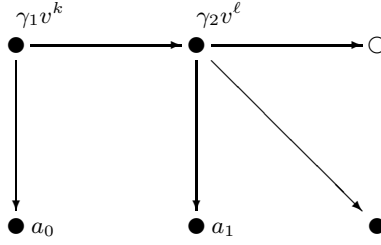

\subsection{Equivariant differential forms on configuration
spaces}\label{ss-6.3}
 Let $\Sigma$ be a manifold with an action of the circle $S^1=\mathbb
R/\mathbb Z$. The infinitesimal action $\mathrm{Lie}(S^1)=\mathbb R
\frac \D{\D t}\to\Gamma(T\Sigma)$ is generated by a vector field
$\vec{v}\in\Gamma(T\Sigma)$, the image of $\frac \D{\D t}$. The
Cartan complex of $S^1$-equivariant forms, computing the equivariant
cohomology with real coefficients, is the differential graded
algebra
\[
\Omega^\bullet_{S^1}(\Sigma)=\Omega^\bullet(\Sigma)^{S^1}[u],
\]
of polynomials in an undetermined $u$ of degree $2$ with
coefficients in the $S^1$-invariant smooth differential forms. The
differential is $\D_{S^1}=\D-u\iota_{\vec{v}}$, where $\D$ is the de
Rham differential and $\iota_{\vec{v}}$ denotes interior
multiplication by $\vec{v}$, extended by $\mathbb R[u]$-linearity.
If $\Sigma$ has an $S^1$-invariant boundary $\partial \Sigma$ and
$j\colon
\partial\Sigma\to \Sigma$ denotes the inclusion map, then the
relative equivariant complex is
\[
\Omega^\bullet_{S^1}(\Sigma,\partial\Sigma) =
\mathrm{Ker}(j^*\colon\Omega^\bullet_{S^1}(\Sigma)\to\Omega^\bullet_{S^1}(\partial\Sigma)).
\]
In the case of the unit disk we have:
\begin{lemma}\label{l-Mei}
Let $\bar D=\{z\in\mathbb C,\,|z|\leq1\}$ be the closed unit disk.
\begin{description}[(iii)]
 \item[(i)]
The equivariant cohomology $H^\bullet_{S^1}(\bar D)$ of $\bar D$ is
the free $\mathbb R[u]$-module generated by the class of
$1\in\Omega^0(\bar D)$.
 \item[(ii)] The relative equivariant cohomology
$H^\bullet_{S^1}(\bar D,\partial \bar D)$ of $(\bar D,\partial\bar
D)$ is the free $\mathbb R[u]$-module generated by the class of
\begin{equation}\label{e-zeromode}
\phi(z,u)=\frac \I{2\pi } \D z\wedge \D\bar z+u(1-|z|^2).
\end{equation}
\end{description}
\end{lemma}
\subsection{The propagator}\label{ss-6.4}
The integrals over configuration spaces defining the
$L_\infty$-morphism are constructed out of a propagator, a
differential 1-form $\omega$ on $\bar D\times \bar D\smallsetminus
\Delta$, with a simple pole on the diagonal $\Delta=\{(z,z),\,z\in
\bar D\}$ and defining the integral kernel of a homotopy contracting
equivariant differential forms to a space of representatives of the
cohomology. The explicit formula of the propagator associated to the
choice of cocycles in Lemma \ref{l-Mei} is given by
\begin{equation}\label{e-propagator}
\omega(z,w)=\frac1{4\pi \I}\left(\D\ln\frac{(z-w)(1-z\bar w)}{(\bar
z-\bar w)(1-\bar z w)}+ z\,\D\bar z-\bar z\,\D z\right).
\end{equation}
\begin{lemma}\label{l-Yuki}
Let $p_i\colon \bar D\times \bar D\to \bar D$ be the projection to
the $i$-th factor, $i=1,2$. The differential form
$\omega\in\Omega^1_{S^1}(\bar D\times \bar D\smallsetminus \Delta)$
has the following properties:
\begin{description}[(iii)]
\item[(i)] Let $j\colon \partial \bar D\times \bar D\to \bar D\times D$ be the
inclusion map. Then $j^*\omega=0$.
\item[(ii)] $\D_{S^1}\omega=-p_1^*\phi$.
\item[(iii)] As $z\to w$, $\omega(z,w)=(2\pi)^{-1}\D\arg(z-w)+$
smooth.
\item[(iv)] As $z$ and $w$ approach a boundary point,
$\omega(z,w)$ converges to the Kontsevich propagator
$\omega_K(x,y)=(2\pi)^{-1}(\D\arg(x-y)-\D\arg(\bar x-y))$ on the
upper half-plane $H_+$ from \cite{Kontsevich2003}. More precisely,
for small $t>0$ let $\varphi_t(x)=z_0 \E^{\I tx}$ be the inclusion
of a neighbourhood of $0\in H_+$ into a neighbourhood of
$z_0\in\partial D$ in $D$. Then $\lim_{t\to
0}(\varphi_t\times\varphi_t)^*\omega=\omega_K$.
\end{description}
\end{lemma}
The proof is a simple computation left to the reader.

\subsection{Weights}\label{ss-Jamie} The weights are integrals of
differential forms over configuration spaces $C^0_{n,m}(D)$ of $n$
points in the unit disk $D=\{z\in\mathbb C\,,\,|z|<1\}$ and $m+1$
cyclically ordered points on its boundary $\partial\bar D$, the
first of which is at $1$:
\begin{eqnarray*}
  C^0_{n,m}(D)&=&\{(z,t)\in D^n\times (\partial \bar D)^m\,,\,z_i\neq
  z_j, (i\neq j),
  \\
  &&  0<\arg(t_1)<\cdots<\arg(t_m)<2\pi\}.
\end{eqnarray*}
The differential forms are obtained from the propagator $\omega$,
see \eqref{e-propagator}, and the form $\phi$, see
\eqref{e-zeromode}. Let $\Gamma\in\mathcal G_{(k_1,\dots,k_n),m}$.
The weight $w_\Gamma$ of $\Gamma$ is
\[
 w_\Gamma=\frac 1{\prod_{i=1}^nk_i!}\int_{C^0_{n,m}(D)}\omega_{\Gamma}
\]
where $\omega\in\Omega^{\bullet}(C^0_{n,m}(D))[u]$ is the
differential form
\[
\omega_{\Gamma}=\prod_{i\in V_1(\Gamma)}\prod_{(i,j)\in
E_b(\Gamma)}\omega(z_i,z_j)\prod_{i\in
V_1(\Gamma)}\phi(z_i,u)^{r_i}.
\]
Here $z_i$ is the coordinate of $z\in C^0_{n,m}(D)$ assigned to the
vertex $i$ of $\Gamma$: to the vertices of the first type we assign
the points in the unit disk and to the vertices of the second type
the points on the boundary. The assignment is uniquely specified by
the ordering of the vertices in $\Gamma$. The number $r_i$ is the
degree of the vertex $i$ plus the number of white vertices connected
to it. The product over $(i,j)$ is over the edges connecting black
vertices to black vertices. For example a point of $C^0_{2,3}(D)$ is given by
coordinates $(z_1,z_2,1,t_1,t_2)$ with $z_i\in D$ and $t_i\in S^1$.
The differential form associated to the graph of Figure
\ref{f-Sheepman}, with degree assignments $k,\ell$  is
\[\pm
\omega_\Gamma=\omega(z_1,1)
\omega(z_1,z_2)\omega(z_2,t_1)\omega(z_2,t_2)\phi(z_1,u)^k\phi(z_2,u)^{\ell+1}.
\]
The signs are tricky. A consistent set of signs may be obtained by
the following procedure. View a multivector field
$\gamma\in\mathfrak g_S[v]$ as a polynomial $\gamma(x,\theta,v)$
whose coefficients are functions on $T^*[1]M=M\times \mathbb
R^d[1]$. Build a function in
$C^\infty((T^*[1]M)^{n+m})[v_1,\dots,v_n]$:
\begin{eqnarray*}
\lefteqn{ g(x^{(1)},\theta^{(1)},v_1,\dots,x^{(\bar m)})}\\
&&=
 \gamma_1(x^{(1)},\theta^{(1)},v_1)
 \cdots
 \gamma_n(x^{(n)},\theta^{(n)},v_n)
 a_0(x^{(\bar 0)})
 \cdots
 a_m(x^{(\bar m)}).
\end{eqnarray*}
Then
\begin{eqnarray*}
 F_n(\gamma;a) &=& (-1)^{|\gamma|m}
 \int_{C^0_{n,m}(D)}i^*_\Delta\circ\exp(\Phi_n)(g)|_{v_1=\dots=v_n=0},\\
 \Phi_n&=&
 \sum_{i\neq k}\omega(z_i,z_k)\sum_{\nu=1}^d\frac{\partial^2}{\partial
\theta^{(i)}_\nu
  \partial x_\nu^{(k)}}
 +\sum_i\phi(z_i,u)\left(\sum_{\nu=1}^d\theta_\nu\frac\partial{\partial
 \theta^{(i)}_\nu}+\frac{\partial}{\partial v_i}\right).
\end{eqnarray*}
The sums over $i$ are from 1 to $n$ and the sum over $k$ is over the
set $\{1,\dots,n,\bar1,\dots,\bar m\}$, with the understanding that
$z_{\bar j}=t_j$. The map $i^*_\Delta$ is the restriction to the
diagonal: its effect is to set all $x^{(i)}$ to be equal to $x$ and
all $\theta^{(i)}$ to be equal to $\theta$. The integrand is then an
element of the tensor product of graded commutative algebras
$\Omega(C^0_{n,m}(D))\otimes C^\infty(T^*[1]M)[u]$. The integral is
defined as $\int(\alpha\otimes \gamma)=(\int\alpha)\gamma$ and the
expansion of the exponential functions gives rise to a finite sum
over graphs.

\subsection{Proof of Proposition \ref{p-Mirelle} on page
\pageref{p-Mirelle}}\label{ss-5last}

A vertex of a directed graph is called disconnected if there is no
edge originating or ending at it.

\begin{lemma}
 Let $\tilde F_n$ be defined as $F_n$ except that the sum
over graphs is restricted to the graphs without disconnected
vertices of the first type. Then
\[
F_{k+n}((hv)^k\cdot\pi^n;f)= \sum_{s=0}^k {k \choose  s} h^s\tilde
F_{k-s+n}((hv)^{k-s}\cdot\pi^n;f).
\]
\end{lemma}

\begin{proof} For each fixed graph $\Gamma_0$ without disconnected
vertices of the first type, we consider all graphs $\Gamma$
contributing to $F_{k+n}$ that reduce to $\Gamma_0$ after removing
all disconnected vertices of the first type. The contribution to
$F_{k+n}$ of such a graph $\Gamma$ is $h^s$ times the contribution
of $\Gamma_0$, where $s$ is the number of disconnected vertices of
the first type. Indeed, each disconnected vertex in a graph $\Gamma$
gives a factor $h$ to $F_\Gamma$ and a factor $\int_D\phi=1$ to the
weight $w_\Gamma$. The proof of the Lemma is complete.
\end{proof}

Let
\[
H_n(\pi,h,f)=\sum_{r=0}^\infty\frac1{r!}\tilde
F_{n+r}((hv)^{r}\cdot\pi^n;f).
\]
In this sum there are finitely many terms since in the absence of
disconnected vertices only derivatives of $h$ can appear and the
number of derivatives is bounded (by $2n$). Therefore $H_n(\pi,h,f)$
is a differential polynomial in $\pi,h,f$. We conclude that
\begin{eqnarray*}
\lefteqn{\sum_{n=0}^\infty\frac1{n!}F_n(\hat\pi^n;f)=
\sum_{n,k=0}^\infty\frac{\epsilon^n}{k!n!}F_{k+n}((hv)^k\pi^n;f)}
\\
&&= \sum_{n,r,s=0}^\infty\frac{\epsilon^n}{r!s!n!}h^s\tilde
F_{n+r}((hv)^r\pi^n;f) =
\E^h\sum_{n=0}^\infty\frac{\epsilon^n}{n!}H_n(\pi,h,f)
\end{eqnarray*}
This concludes the proof of Proposition \ref{p-Mirelle}.

\section{Equivariant differential forms on configuration spaces
and Stokes theorem}   \label{s-6}
\subsection{Configuration spaces and their compactifications} \label{ss-6.1}
 We consider three types of configuration spaces of
points, the first two appearing in \cite{Kontsevich2003}.

\noindent{(i) \emph{Configuration spaces of points in the plane}.}
Let $\mathrm{Conf}_n(\mathbb C)=\{z\in\mathbb C^n\,,\, z_i\neq z_j,
(i\neq j)\}$, $n\geq 2$. The three-dimensional real Lie group $G_3$
of affine transformations $w\mapsto aw +b,a>0,b\in\mathbb C$ acts
freely on the manifold $\mathrm{Conf}_n(\mathbb C)$. We set
$C_n(\mathbb C)=\mathrm{Conf}_n(\mathbb C)/G_3$ ($n\geq2$). It is a
smooth manifold of dimension $2n-3$. We fix the orientation defined
by the volume form $\D\varphi_2\wedge
\bigwedge_{j\geq3}\D\mathrm{Re}(z_j)\wedge \D\mathrm{Im}(z_j)$, with
the choice of representatives with $z_1=0,z_2=\E^{\I\varphi_2}$.

\noindent{(ii) \emph{Configuration spaces of points in the upper
half-plane}.} Let $H_+=\{z\in\mathbb C\,,\,\mathrm{Im}(z)> 0\}$ be
the upper half-plane. Let $\mathrm{Conf}_{n,m}(H_+)=\{(z,x)\in
H_+^n\times\mathbb R^m,\, z_i\neq z_j, (i\neq j), t_1<\cdots<t_m\}$,
$2n+m\geq2$. The two-dimensional real Lie group $G_2$ of affine
transformations $w\mapsto aw +b,a>0,b\in\mathbb R$ acts freely on
the manifold $\mathrm{Conf}_{n,m}(H_+)$. We set
$C_{n,m}(H_+)=\mathrm{Conf}_{n,m}(H_+)/G_2$ ($2n+m\geq2$). It is a
smooth manifold of dimension $2n+m-2$.  If $n\geq1$ we fix the
orientation by choosing representatives with $z_1=i$ and taking the
volume form $\D t_1\wedge\cdots\wedge \D
t_m\wedge\bigwedge_{j\geq2}\D\mathrm{Re}(z_j)\wedge
\D\mathrm{Im}(z_j)$. If $m\geq2$ we fix the orientation defined by
the volume form $(-1)^m\D t_2\wedge\dots\wedge \D t_{m-1}\wedge
\bigwedge_{j\geq1}\D\mathrm{Re}(z_j)\wedge \D\mathrm{Im}(z_j)$, with
the choice of representatives with $t_1=0,t_m=1$. If $m\geq2$ and
$n\geq1$ it is easy to check that the two orientations coincide.

\noindent{(iii) \emph{Configuration spaces of points in the disk}.}
 Let $D=\{z\in\mathbb C\,,\,|z|<1\}$ be the
unit disk, $S^1=\partial \bar D$ the unit circle. Let
$C_{n,m+1}(D)=\{(z,x)\in D^n\times(S^1)^{m+1},\, z_i\neq z_j, (i\neq
j), \arg(t_0)<\cdots<\arg(t_m)<\arg(t_0)+2\pi\}$, $m\geq0$. The
circle group acts freely on $C_{n,m+1}(D)$ by rotations. We do not
take a quotient here, since the differential forms we will introduce
are not basic, and work equivariantly instead. Instead of the
quotient we consider the section $C^0_{n,m}(D)=\{(z,x)\in
C_{n,m+1}(D)\,,\, t_0=1\}$, $(m\geq1)$. It is a smooth manifold of
dimension $2n+m$. The orientation of $C_{n,m+1}(D)$ is defined by
$\D\arg(t_0)\wedge\cdots\wedge
\D\arg(t_m)\wedge\bigwedge_{j=1}^n\D\mathrm{Re}(z_j)\wedge
\D\mathrm{Im}(z_j)$. The orientation of $C^0_{n,m}(D)$ is defined by
$\D\arg(t_1)\wedge\cdots\wedge
\D\arg(t_m)\wedge\bigwedge_{j=1}^n\D\mathrm{Re}(z_j)\wedge
\D\mathrm{Im}(z_j)$.

As in \cite{Kontsevich2003}, compactifications $\bar C_{n}(\mathbb
C)$, $\bar C_{n,m}(H_+)$, $\bar C_{n,m+1}(D)$, $\bar C^0_{n,m}(D)$
of these spaces as manifolds with corners are important. Their
construction is the same as in \cite{Kontsevich2003}. Roughly
speaking, one adds strata of codimension $1$ corresponding to
limiting configurations in which a group of points collapses to a
point, possibly on the boundary, in such a way that within the group
the relative position after rescaling remains fixed. Higher
codimension strata correspond to collapses of several groups of
points possibly within each other. The main point is that the Stokes
theorem applies for smooth top differential forms on manifold with
corners, and for this only codimension 1 strata are important.

 Let us describe the codimension 1 strata of $C^0_{n,m}(D)$.

\noindent{\emph{Strata of type I}.} These are strata where a subset
$A$ of $n'\geq 2$ out of $n$ points $z_i$ in the interior of the
disk collapse at a point in the interior of the disk, the relative
position of the collapsing points is described by a configuration on
the plane and the remaining points and the point of collapse are
given by a configuration on the disk. This stratum is thus
\begin{equation}\label{e-Mimi}
\partial_A\bar C^0_{n,m}(D)\simeq \bar C_{n'}(\mathbb C)\times \bar C^0_{n-n'+1,m}(D).
\end{equation}

\noindent{\emph{Strata of type II}.} These are strata where a subset
$A$ of $n'$ out of $n$ points $z_i$ and a subset $B$ of the $m$
points $t_i$ collapse at a point on the boundary of the disk
($2n'+m'\geq 2$). The relative position of the collapsing points is
described by a configuration on the upper half-plane and the
remaining points and the point of collapse are given by a
configuration on the disk. This stratum is thus
\begin{equation}\label{Joe}
\partial_{A,B}\bar C^0_{n,m}(D)\simeq \bar C_{n',m'}(H_+)\times \bar C^0_{n-n',m-m'+1}(D).
\end{equation}

\subsection{Forgetting the base point and cyclic shifts} \label{ss-6.2}
 Let $j_0\colon C^0_{n,m}(D)\to C_{n,m}(D)$ be the map
$(z,1,t_1,\dots,t_m)\mapsto (z,t_1,\dots,t_m)$ forgetting the base
point $t_0=1$. It is an orientation preserving open embedding.

The cyclic shift $\lambda\colon C^0_{n,m}(D)\to C^0_{n,m}(D)$ is the
map
\[
\lambda\colon(z_1,\dots,z_n,1,t_1,\dots,t_m)\mapsto
(z_1,\dots,z_n,1,t_m,t_1,\dots,t_{m-1}).
\]
It is a diffeomorphism preserving the orientation if $m$ is odd and
reversing the orientation if $m$ is even. The following fact is then
easily checked.

\begin{lemma}\label{l-Isabel}
The collection of maps $j_k=j_0\circ\lambda^{\circ k}$,
$k=0,\dots,m-1$ defines an embedding $j\colon
C^0_{n,m}(D)\sqcup\cdots\sqcup C^0_{n,m}(D)\to C_{n,m}(D)$ with
dense image. The restriction of $j$ to the $k$th copy of
$C_{n,m}(D)$ multiplies the orientation by $(-1)^{(m-1)k}$.
\end{lemma}

\subsection{Proof of Theorem \ref{t-Grace} on page \pageref{t-Grace}}\label{ss-6.6}
The proof uses the Stokes theorem as in \cite{Kontsevich2003}. The
new features are: (i) the differential forms in the integrand are
not closed and (ii) an equivariant version of the Stokes theorem is
used.

We first compute the differential of the differential form
associated to a graph $\Gamma$.
\begin{lemma}\label{l-Abe}
Let $\partial_e\Gamma$ be the graph obtained from $\Gamma$ by adding
a new white vertex $*$ and replacing the black-to-black edge $e\in
E_b(\Gamma)$ by an edge originating at the same vertex as $e$ but
ending at $*$. Then
\[
\D_{S^1}\omega_\Gamma=\sum_{e\in E_b(\Gamma)}(-1)^{\sharp e}
\omega_{\partial_e\Gamma},
\]
where $\sharp e=k$ if $e=e_k$ and $e_1,\dots,e_N$ are the edges of
$\Gamma$ in the ordering specified by the ordering of the vertices
and of the edges at each vertex.
\end{lemma}
\begin{proof}
This follows from the fact that $\D_{S^1}$ is a derivation of degree
1 of the algebra of equivariant forms and Lemma \ref{l-Yuki}, (ii).
\end{proof}
The next Lemma is an equivariant version of the Stokes theorem.
\begin{lemma}\label{l-Gotanda}
Let $\omega\in\Omega_{S^1}^{\bullet}(\bar C_{n,m+1}(D))$. Denote
also by $\omega$ its restriction to $\bar C^0_{n,m}(D)\subset \bar
C_{n,m+1}(D)$ embedded as the subspace where $t_0=1$ and to the
codimension 1 strata $\partial_iC^0_{n,m}(D)$ of $C^0_{n,m}(D)$.
Then
\[
\int_{\bar C^0_{n,m}(D)}\D_{S^1}\omega=\sum_i\int_{\partial_i\bar
C^0_{n,m}(D)}\omega-u\int_{\bar C_{n,m+1}(D)}\omega.
\]
\end{lemma}
\begin{proof}
Write $\D_{S^1}=\D-u\iota_{\vec{v}}$. For $u=0$ the claim is just
the Stokes theorem for manifolds with corners. Let us compare the
coefficients of $u$. The action map restricts to a diffeomorphism
$f\colon S^1\times \bar C^0_{n,m}(D)\to \bar C_{n,m+1}(D)$. Since
$\omega$ is $S^1$-invariant, $\iota_{\vec{v}}\omega$ is also
invariant and we have $f^*\omega=1\otimes\omega+\D
t\otimes\iota_{\vec{v}}\omega\in\Omega(S^1)\otimes\Omega(\bar
C^0_{n,m}(D))\subset\Omega(S^1\times\bar C^0_{n,m}(D))$, where $t$
is the coordinate on the circle $S^1=\mathbb R/\mathbb Z$. Thus
\[
\int_{\bar C_{n,m+1}(D)}\omega=\int_{S^1\times \bar C^0_{n,m}(D)}\D
t\otimes \iota_{\vec{v}}\omega=\int_{\bar
C^0_{n,m}(D)}\iota_{\vec{v}}\omega.
\]
\end{proof}

Finally we use Lemma \ref{l-Isabel} to reduce the integral over
$\bar C_{n,m+1}(D)$ to integrals over $\bar C_{n,m+1}^0(D)$. We
obtain:
\begin{lemma}\label{l-Marumaki}
Let $\omega\in\Omega_{S^1}^{\bullet}(\bar C_{n,m+1}(D))$ and let
$j_k$ be the maps defined in Lemma \ref{l-Isabel}. Then
\[
\int_{\bar C_{n,m+1}(D)}\omega=\sum_{k=0}^{m}(-1)^{mk}\int_{\bar
C^0_{n,m+1}(D)}j_k^*\omega.
\]
\end{lemma}

We can now complete the proof of Theorem \ref{t-Grace}. We first
prove the identity \eqref{e-Miranda}, starting from the right-hand
side. Suppose that $a=(a_0,\dots,a_m)\in C_{-m}(A)$,
$\gamma=\gamma_1\cdots\gamma_n$, with
$\gamma_i\in\Gamma(\wedge^{k_i}\mathit{TM})$. It is convenient to
identify $\Gamma(\wedge \mathit{TM})$ with
$C^\infty(M)[\theta_1,\dots,\theta_n]$ where $\theta_i$ are
anticommuting variables, so that
$\Div_\Omega=\sum\partial^2/\partial t_i\partial\theta_i$. It
follows that for any $\Gamma\in \mathcal G_{\mathbf k,m}$,
$\Div_\Omega F_\Gamma(\gamma;a)$ can be written as a sum (with
signs) of terms $F_{\Gamma'}(\gamma;a)$, where $\Gamma'$ is obtained
from $\Gamma$ by identifying a white vertex with a black vertex and
coloring it black. Some of these graphs $\Gamma'$ have an edge
connecting a vertex to itself and contribute to
$F_n(\delta_\Omega\gamma;a)$.  The remaining ones yield, in the
notation of Lemma \ref{l-Abe}:
\[
\Div_\Omega F_n(\gamma;a)-F_n(\delta_\Omega\gamma;a)=
\sum_{(\Gamma,e)}(-1)^{\sharp
e}w_{\partial_e\Gamma}F_{\Gamma}(\gamma;a).
\]
The summation is over pairs $(\Gamma,e)$ where $\Gamma\in\mathcal
G_{\mathbf k,m}$ and $e\in E_b(\Gamma)$ is a black-to-black edge. By
Lemma \ref{l-Gotanda} and \ref{l-Marumaki},
\[
 \sum_{e\in E_b(\Gamma)}
 (-1)^{\sharp e}w_{\partial_e\Gamma}
 =\sum_i\int_{\partial_iC^0_{n,m}(D)}\omega_\Gamma
 -u \sum_{k=0}^m(-1)^{km}
 \int_{\bar C^0_{n,m+1}}j_k^*\omega_\Gamma.
\]
The second term on the right-hand side, containing the sum over
cyclic permutations, gives rise to $F_{n+1}(\gamma;Ba)$. The first
term is treated as in \cite{Kontsevich2003}: the strata of type I
(see Section \ref{ss-6.1}) give zero by Kontsevich's lemma (see
\cite{Kontsevich2003}, Theorem 6.5) unless the number $n'$ of
collapsing interior points is 2. The sum over graphs contributes
then to the term with the Schouten bracket $[\gamma_i,\gamma_j]$ in
\eqref{e-Miranda}. The strata of type II such that $n-k>0$ interior
points approach the boundary give rise to the term containing the
components of the Kontsevich $L_\infty$-morphism $U_{n-k}$. Finally
the strata of type II in which only boundary points collapse give
the term with Hochschild differential $F_{n-1}(\gamma;ba)$. This
proves \eqref{e-Miranda}.

Property (i) is clear: $F_0$ is a sum over graphs with vertices of
the second type only. These graphs have no edges. Thus the only case
for which the weight does not vanish is when the configuration space
is $0$-dimensional, namely when there is only one vertex. Property
(ii) is checked by an explicit calculation of the weight. The only
graphs with a non-trivial weight have edges connecting the vertex of
the first type with white vertices or to vertices of the second
type. There must be at least $p$ edges otherwise the weight vanishes
for dimensional reasons. In this case, i.e. if $k\geq p$, the
integral computing the weight is
\begin{equation}
w_\Gamma=\int \phi(z,u)^{\ell+k-p}\omega(z,t_1)\cdots\omega(z,t_p),
\end{equation}
with integration over $z\in D$, $t_i\in S^1$,
$0<\arg(t_1)<\cdots<\arg(t_p)<2\pi$. The integral of the product of
the 1-forms $\omega$ is a function of $z$ that is independent of
$z$, as is easily checked by differentiating with respect to $z$,
using the Stokes theorem and the boundary conditions of $\omega$.
Thus it can be computed for $z=0$. Since
$\omega(0,t_i)=\frac1{2\pi}\D \arg(t_i)$ the integral is $1/p!$. The
remaining integral over $z$ can then be performed. Set
$\ell+k-p=s+1$. This power must be positive otherwise the integral
vanishes for dimensional reasons.
\[
\int_D \phi(z,u)^{s+1}=\frac\I{2\pi}(s+1)u^s\int_D(1-|z|^2)^s\D
z\wedge\D\bar z=u^s,\quad s\geq 0,
\]
and we obtain $w_\Gamma=u^s/p!$. We turn to Property (iii). The
equivariance under linear coordinate transformations is implicit in
the construction. The graphs contributing to $F_n(\gamma_1\cdots;a)$
for linear $\gamma_1$ are of two types: either the vertex associated
with $\gamma_1$ has exactly one ingoing and one outgoing edge or it
has an outgoing edge pointing to a white vertex and there are no
incoming edges. The graphs of the second type contribute to
$\gamma_1\wedge F_{n-1}(\cdots;a)$, since their weight factorize as
$1=\int_D\phi$ times the weight of the graphs obtained by omitting
the vertex associated to $\gamma_1$ and the white vertex connected
to it. The claim then follows from the following vanishing lemma.
\begin{lemma}
\begin{enumerate}
\item[(i)] For all $z,z'\in \bar D$,
$\int_{w\in D}\omega(z,w)\omega(w,z')=0$.
\item[(ii)] For all $z\in\bar D$, $\int_{w\in
D}\omega(z,w)\phi(w,u)=0$.
\end{enumerate}
\end{lemma}
\begin{proof} (i)
We reduce the first claim to the second: consider the integral
\[
I(z,z')=\int_{w_1,w_2\in
D}\D(\omega(z,w_1)\omega(w_1,w_2)\omega(w_2,z')).
\]
On one hand, $I(z,z')$ can be evaluated by using Stokes's theorem,
giving three terms all equal up to sign to the integral appearing in
(i). On the other side, the differential can be evaluated explicitly
giving
\[
I(z,z')=-\int_{w_1,w_2\in D}\omega(z,w_1)\omega(w_1,w_2)\phi(w_2,0).
\]
The integral over $w_2$ vanishes if (ii) holds. The proof of (ii) is
an elementary computation that uses the explicit expression of
$\omega$ and $\phi$. Alternatively, one shows that $\int_{w\in
D}\omega(z,w)\phi(w,u)$ is a closed 1-form on the disk that vanishes
on the boundary, is invariant under rotations and odd under diameter
reflections. Therefore it vanishes. We leave the details to the
reader.
 \end{proof}
\bibliographystyle{plain}
\bibliography{Cyclic8}

\end{document}